\documentclass[12pt]{iopart}
\usepackage{iopams}  
\usepackage{epsf}  
\begin{document}

\title[
]
{Model wave functions and dynamic correlations in
light-medium nuclei
}

\author
{E Buend\'{\i}a\dag , F J G\'alvez\dag ,
J Praena\dag\ and A Sarsa\ddag
\footnote[3]{To
whom correspondence should be addressed (fa1sarua@uco.es)},
}

\address
{\dag\ Departamento de F\'{\i}sica Moderna, Facultad de Ciencias, 
Universidad de Granada,  E-18071 Granada, Spain}
\address
{\ddag\  Departamento de F\'{\i}sica, Campus de Rabanales, Edificio C2,
Universidad de C\'ordoba, E-14071 C\'ordoba, Spain
}

\begin{abstract}
The nuclei $^4$He, $^8$Be, $^{12}$C and $^{16}$O have been studied
starting from nucleon-nucleon interactions of $v_4$ type.  The wave
function is built as the product of three terms, a Jastrow correlation
factor, a linear correlation factor and a model wave function.  The
correlation factors account for both the short range repulsive and the
spin and isospin dependence of the nuclear potential.  The model wave
function is antisymmetric and has the values of the angular momentum
and parity of the state under description.  For the model wave function
we have used two different schemes.  The first one is based on a
Harmonic Oscillator shell model with and without deformation, and the
second one is based on the Margenau-Brink model of alpha clustering.
Projection operators of parity and total angular momentum are used.
The performance of these two models is studied and compared
systematically.  Wave functions for the ground state and some members
of its rotational band and some other bound states of these nuclei
have been obtained.  Binding energies, root mean square radius and the
expectation value of the kinetic energy and the different channels of
the nuclear interactions and the one-- and two-- body densities are
reported.  The two different model wave functions and the effects of
the different nucleon-nucleon correlations have been evaluated on
those quantities.  All the results here presented have been obtained
by using the Variational Monte Carlo method.
\end{abstract}


\maketitle

\section{Introduction}
\label{sec.intro}

The use of explicitly correlated trial wave functions to study nuclear
bound states constitutes an efficient and compact way to include the
complex dynamic mechanisms induced by the nuclear potential in the
nuclear structure.
Short range correlations along with a model wave function is a scheme widely
used as starting point when realistic or semi-realistic interactions
are considered.  Usually, the model wave function is a Slater
determinant built with orbitals obtained from a given mean field.
By including in the ansatz correlation factors, one
tries to take care of the deficiencies of this model wave function,
specially the dependence of the nuclear force on the state of the
nucleon pair and the behaviour at short internucleon distances.

Jastrow central correlation factors \cite{jast55} have shown to be an
appropriate way of handling the strongly repulsive short range
core of the nuclear force in the operator independent channel.  This
functional form has been extended to the spin and isospin dependent
channels including Jastrow type operatorial correlation factors. 
In this way the model wave function is corrected, and the two nucleons
dynamic is adapted to the behaviour of the different channels of the
interaction. A major drawback of this ansatz is the enormous
computational difficulties posed by this wave function, mainly due to the
presence of operatorial correlations.  In fact this correlated
wave function has been applied so far only to light nuclei
\cite{pvw02}, or by using some approximations for
heavier nuclear systems  \cite{arias96,morales02}.

In order to overcome these problems, simpler trial wave functions with
a linear state dependent correlation factor \cite{guardiola96}, and a central
Jastrow term times a linear state dependent correlation factor
\cite{bgmnppw98} have been proposed. 
With them medium nuclei have been studied obtaining good results 
\cite{guardiola96,bgmnppw98,bgmnpw99,bgps00,bgps01,bgps02}. 
In this way, the short range effects, at least in the Wigner channel,
are adequately described by means of the central Jastrow factor.
The linear term can be obtained from the Coupled Cluster method
including only two--body translationally and rotationally invariant
excitations \cite{guardiola96,bfbbg90}, formally equivalent to a Configuration
Interaction expansion.
Thus, the use of a linear correlation factor can be viewed either as a
linear approximation to the Jastrow operatorial factor or as 
Configuration Interaction expansion of the trial wave function.

The variational freedom given by the model part of the wave function
can be exploited to include medium and long range effects induced by
the nuclear interaction.  In general, it is difficult to describe
these effects by means of a correlation factor.  These are typical
many body effects that are usually treated in nuclear physics by
using, for instance, a deformed mean field.  An important example of
these kind of effects is the formation of alpha clusters or other type
of nucleon groupings.  In principle one could describe these effects
in terms of correlation factors by, for example, truncating the
Coupled Cluster exp(S) wave function at higher orders, considering in
this way three--, four--, and larger $n$--body excitations.  However,
and within the scheme of this work, it is simpler to incorporate long
range effects in the model wave function, by making use of the different
models developed for effective interactions.

Therefore the different components of the trial wave function are tailored
to account for different correlation mechanisms.
The aim of this work is to use these wave functions to study 
the ground and some low lying excited states of the
$^4$He, $^{8}$Be, $^{12}$C and $^{16}$O nuclei and to analyse the interplay
between the different dynamical effects included in the wave function.
The ansatz consists of three factors: a state independent Jastrow
correlation factor, a linear state dependent factor and a model wave function.
Two different model wave functions are employed.  The first
one is based on a shell model built from a deformed Harmonic
Oscillator mean field while the second one is based on the
$\alpha$-cluster Margenau-Brink model. 
The determination of the optimal variational wave function and the
calculation of the different nuclear properties is carried out by means
of the Variational Monte Carlo method.

The structure of this work is as follows.  In section \ref{sec.trialfunc}
we show in detail the variational trial wave function used.  Section
\ref{sec.vmc} is devoted to the technical aspects involved in the
calculation of the different properties.  The results are presented
and discussed in section \ref{sec.results}. 
Finally, section \ref{sec.conclusions} summarizes and gives the
conclusions of this work.

\section{Trial wave function}
\label{sec.trialfunc}

The variational trial wave functions used are the product of a
correlation factor, which includes a central Jastrow correlation
factor and a linear state-dependent correlation factor, and a model
wave function
\begin{equation}
\label{trialwf}
\Psi^{\pm}_{JKM}(1,\ldots,A)=
F_{\cal J}(1,\ldots,A) \, F_{\cal L}(1,\ldots,A) \Phi^{\pm}_{JKM}(1,\ldots,A)
\end{equation}

The function $F_{\cal J}$ is the  central Jastrow factor that depends only on
the distance between pairs of nucleons
\begin{equation}
F_{\cal J}(1,\ldots,A)=\prod_{i<j}^A f(r_{ij})
\end{equation}
while  $F_{\cal L}$ is the linear state-dependent correlation factor defined as
\begin{equation}
F_{\cal L}(1,\ldots,A)=\sum_{i<j}^A g(i,j).  
\end{equation}
where the function $g(i,j)$ depends on the radial and intrinsic
degrees of freedom of the particles, $i$ and $j$
\begin{equation}
g(i,j)=g^{(1)}(r_{ij})+\sum_{k=2}^{4}g^{(k)}(r_{ij}){\bf P}^{(k)}(i,j),
\end{equation}
and ${\bf P}^{(k)}(i,j)$, with $k=2,3,4$, are the exchange spin, isospin and 
spin--isospin operators, respectively. 
This is the only part of the trial wave function where state dependent
correlations are present explicitly.

The radial correlation functions $g^{(k)}(r)$, $k=1,..,4$, and $f(r)$
are parameterized as a linear combination of Gaussian functions
\begin{equation}
g^{(k)}(r)=\sum_{m=0}^{M} a_m^{(k)}~e^{-b_m^2 r^2},~~~
f(r)=1+\sum_{n=1}^{N} c_n~e^{-d_n^2 r^2}.  
\label{expansion}
\end{equation}

The structure of the state dependent correlation factor is the
same as that of the nuclear potential.  We have used two different
$v_4$ type nuclear forces with the same operatorial dependence as the
linear correlation factor; the BB1 Brink-Boeker \cite{BrBo-67} 
and a modified S3 Afnan-Tang interaction \cite{AfTa-68}.
The first one was fitted by using non
correlated wave functions.  The second one is a semi-realistic
interaction fixed by using explicitly correlated wave functions to
reproduce the $s$-wave scattering data up to about 60 MeV for the
alpha particle.  Some of the parameters of this interaction were
modified by Guardiola \cite{guardiola81} to include the repulsion in
the triplet states, necessary to an adequate description of nuclei heavier
than the $\alpha$ particle.
Both potentials have been extensively employed in previous works.

For the model wave function, $\Phi^{\pm}_{JKM}$, we have considered
two possibilities, a shell model wave function built from Cartesian
Harmonic oscillator orbitals, and a wave function including the alpha
clustering effect by means of the Margenau-Brink model. This model was
devised by Margenau \cite{marge41} as a simpler alternative to the
Wheeler model \cite{wheeler37}, and it was further developed by Brink
\cite{brink66}.  
Within this scheme, nuclear bound states are
described as a composite of $\alpha$ particles centered around some
given fixed positions, like the atoms in a molecule. Different
arrangements of the $^4$He clusters give rise to different geometries
and states. The model function is taken to be antisymmetric with
respect to nucleon exchange.

We study $A=4 n$, $n=1,2,3,4$  nuclei with $N=Z$. The model wave
function based on the  Margenau-Brink model can be written as 
\begin{equation}
\Phi_{\vec{{\bf C}}}(1,2,\ldots, A)= 
{\cal  A}
\left\{\prod_{k=1}^n\xi_{\vec{c}_{k}}(4k-3,4k-2,4k-1,4k)\right\}
\end{equation}
where $\vec{\bf C}\equiv\left\{\vec{c}_k\right\}_{k=1}^n$ is a set of
centers of the $\alpha$-cluster, and the operator ${\cal A}$ is the
antisymmetrizer of $A$ particles.  The $\xi_{\vec{c}_{k}}$ functions
can be any approximated wave function for the $^4$He nucleus centered
at $\vec{c}_k$. In this work we have taken these functions to be of
the form
\begin{equation}
\fl
\xi_{\vec{c}_{k}}(1,2,3,4)=\frac{1}{\sqrt{4!}}
    \left|\begin{array}{llll}
    \phi_{\vec{c}_k}(\beta;\vec{r_1})\eta_1(1) &
    \phi_{\vec{c}_k}(\beta;\vec{r_1})\eta_2(1) & \ldots &
    \phi_{\vec{c}_k}(\beta;\vec{r_1})\eta_4(1)\\
    \vdots & \vdots & \ddots & \vdots \\
    \phi_{\vec{c}_k}(\beta;\vec{r_4})\eta_1(4) &
    \phi_{\vec{c}_k}(\beta;\vec{r_4})\eta_2(4) & \ldots &
    \phi_{\vec{c}_k}(\beta;\vec{r_4})\eta_4(4)
\end{array}\right|
\end{equation}
where $\eta_k(i)$ stands for the four possible spin-isospin states of
the nucleon, and the single particle wave function is taken to be
a $s$-wave harmonic oscillator orbital centered at $\vec{c}_k$
\begin{equation}
\phi_{\vec{c}_k}(\beta;\vec{r})=
\left(
\frac{\beta^2}{\pi}
\right)^{3/4}
e^{-\frac{1}{2}\beta^2(\vec{r}-\vec{c}_k)^2}  
\end{equation}

It is straightforward to check that the full model wave function
becomes a Slater determinant, that allows one to apply the machinery
developed to deal with determinants in Variational Monte Carlo calculations.

Regardless the actual values of the centers, this function has
total spin equal to zero. By using $\Phi_{\vec{\bf C}}$ and
$\Phi_{-\vec{\bf C}}$ (the same function with centers at opposite positions)
we can build eigenfunctions of the parity operator
\begin{equation}
\Phi^{\pm}_{\vec{\bf C}}(1,\ldots, A)=\Phi_{\vec{\bf
    C}}(1,\ldots, A) \pm \Phi_{-\vec{\bf C}}(1,\ldots, A).  
\end{equation}

In general these functions have not a definite value of the total angular
momentum. In order to get model wave functions that are eigenfunctions
of the angular momentum operator, the Peierls-Yoccoz projection 
operators  \cite{peyo57}, ${\cal P}_{MK}^J$, are used. 
In this way we obtain the model wave function 
\begin{eqnarray}
\fl
\Phi^{\pm}_{JKM;\vec{\bf C}}(1,\ldots, A) &=&
{\cal P}_{MK}^J~ \Phi^{\pm}_{\vec{\bf C}}(1,\ldots, A) 
\nonumber
\\
&=& \frac{2J+1}{8\pi^2}\int d\Theta 
{\cal D}_{MK}^{J*}(\Theta){\bf R}(\Theta) 
\Phi^{\pm}_{\vec{\bf C}}(1,\ldots, A)
\label{projection}
\end{eqnarray}
where ${\bf R}(\Theta)$ is the rotation operator, ${\cal
D}_{MK}^{J*}(\Theta)$ the rotation matrix and $\Theta$ represents the
Euler angles. The quantum number $J$ gives the total angular momentum,
$K $ is the projection along the nuclear $z$ axis and $M $ is the
projection along the $Z$ axis of the laboratory fixed frame. Therefore
the projection is carried out by rotating the intrinsic state and
integrating over all angles weighted by the rotation matrix.  The
projection operation is done only for the model part of the trial wave
function because both the Jastrow and the linear correlation factors commute
with the projection operator.

Within this scheme a rotational band is associated to each value of
$K$. The states in the band correspond to all the possible values of
$J$ and parity allowed by the symmetry of the spatial arrangements of
the $\alpha$ clusters \cite{brink66,ahkt72,iklp79,dude96}.  For
$^{8}$Be only a linear configuration can be built within this model
with the two alpha clusters separated by a distance $d$. This
corresponds to the point symmetry group D$_{\infty h}$. For $^{12}$C
one can build different arrangements of the $\alpha$ cluster giving
rise to different geometries. In a non-correlated calculation the
lowest energy is obtained from an equilateral triangle having D$_{3
h}$ as symmetry group. This configuration will be used here to
describe, when $K=0$, the $0^+$ ground state of this nucleus and
its rotational band, as well the $J^{\pi} = 3^-$ excited state, with
$K=3$. We shall use a linear configuration of the three alpha
particles to study the first excited $0^+$ state whose excitation
energy is $7.65$ MeV.  Finally, the $0^+$ ground state and the first
excited state with $J^{\pi} = 3^-$ of $^{16}$O will be described with
the $\alpha$ particles lying at the vertices of a regular tetrahedron
and therefore with T$_d$ as symmetry group.  For this nucleus we shall
also consider the linear configuration of the alpha particles.  In
table \ref{table1} we summarize the different configurations studied
here with the corresponding $K$ and $J$ values and the parity of the
states.

\Table{
\label{table1}
Symmetry point group of the different
spatial arrangements of the $\alpha$ clusters and the limiting 
Harmonic Oscillator shell model configurations (HO).
The different rotational states that can be obtained are also given.

}
\br
         &                 & $(K:J^{\pi})$  & HO\\
\mr
$^8$Be   & $D_{\infty h}$  & $(0: 0^+, 2^+,4^+)$ 
& $[(0,0,0)^4(0,0,1)^4]$               \\
$^{12}$C & $D_{3h}$        & $(0: 0^+, 2^+,4^+)$ 
& $[(0,0,0)^4(1,0,0)^4(0,1,0)^4]$       \\
         &                 & $(3: 3^-)$ 
&                                       \\
$^{12}$C &  $D_{\infty h}$ & $(0: 0^+, 2^+,4^+)$ 
& $[(0,0,0)^4(0,0,1)^4(0,0,2)^4]$        \\
$^{16}$O & $T_d$           & $(0: 0^+)$ &
$[(0,0,0)^4(1,0,0)^4(0,1,0)^4(0,0,1)^4]$  \\
         &                 & $(2: 3^-)$ &\\ 
$^{16}$O & $D_{\infty h}$  & $(0: 0^+)$ 
& $[(0,0,0)^4(0,0,1)^4(0,0,2)^4(0,0,3)^4]$  \\
\br
\endTable

When the distances between the different alpha clusters are taken to be
zero some of the Margenau-Brink model functions tend to shell model
wave functions. In table \ref{table1} we give the limiting Harmonic
Oscillator shell model configurations of the corresponding Margenau-Brink 
wave functions for zero intercluster distance. 
Any of these  wave functions is written as a Slater determinant with not 
well defined value of the angular momentum, except for the tetrahedron in
$^{16}$O.  
For states with axial symmetry
we have considered deformed harmonic oscillator wave functions.  By
including axial symmetry in the model wave functions a noticeable
improvement in the binding energy is obtained with respect to the
spherical shell model approximation.

\section{VMC calculation of the matrix elements}
\label{sec.vmc}

The VMC evaluation of the  expectation values involved in the
determination of the energy and other properties presents several
differences with respect to standard algorithms \cite{kawh86}.  The
reason for that lies in the presence in the trial wave function of
both, the state dependent correlations (by means of $F_{\cal L}$), and the
projection operator ${\cal P}_{MK}^J$. The technical problems induced
by each one of these elements can be treated independently. The
spin-isospin dependence in the trial wave function is the main
source of difficulties and the ultimate reason why the nuclear
problem is more complex, from a computational point of view, than
other non-relativistic many body systems.

To deal with  angular momentum projection we use equation (\ref{projection})
to write the expectation value of the Hamiltonian, $H$, as
\begin{equation}
\label{expectation}
\frac{\langle\Psi^{\pm}_{JKM}|H|\Psi^{\pm}_{JKM}\rangle}
{\langle\Psi^{\pm}_{JKM}|\Psi^{\pm}_{JKM}\rangle}= 
\frac{\int d\Omega \,{\cal D}_{KK}^{J*}(\Omega)
\langle\Phi^{\pm}|F_{\cal J}F_{\cal L}~H~F_{\cal J}F_{\cal L}{\bf R}(\Omega)|\Phi^{\pm}\rangle}
{\int d\Omega \, {\cal D}_{KK}^{J*}(\Omega)
\langle\Phi^{\pm}|F_{\cal J}F_{\cal L}~F_{\cal J}F_{\cal L}{\bf R}(\Omega)|\Phi^{\pm}\rangle}
\end{equation}

In this equation the transformation properties of the rotation
matrices and the rotational invariance of the correlation factors have
been taken into account \cite{brink66}.  
Note that this is an integral over the particle degrees of freedom,
intrinsic and spatial considered in the bracket, and the Euler
angles of the rotation. The latter can be further simplified for nuclei
with axial symmetry.  This multi dimensional integration is performed
here by using a Monte Carlo quadrature with the Metropolis
algorithm. The probability distribution function used will be
discussed below. Finally, it is worth to point out that the action of
the rotation operator over a spin-isospin saturated Slater determinant
is another spin-isospin saturated Slater determinant. This property is
important for our treatment of the state-dependent correlations.

To carry out the spin-isospin integral in the expectation value of
the Hamiltonian, equation (\ref{expectation}), we use the following complete set
\begin{equation}
|{\cal R},\Xi\rangle =
|\vec{r}_{1}\eta_{1}
\rangle \, 
|\vec{r}_{2}\eta_{2}
\rangle \ldots
|\vec{r}_{A}\eta_{A}
\rangle 
\end{equation}
where $\vec{r}_i$ is the spatial coordinate of the i-particle and
$\eta_i$ represents its spin-isospin variables.  ${\cal R}$ and $\Xi$
denote the spatial coordinates and spin-isospin components,
respectively, of the whole system.

Thus the expectation value in equation (\ref{expectation}), can be written as
\begin{equation}
\fl
\langle\Phi^{\pm}|F_{\cal J}F_{\cal L}~H~F_{\cal J}F_{\cal L}{\bf R}(\Omega)|\Phi^{\pm}\rangle =
\sum \int d{\cal R}~
\langle \Phi^{\pm}|F_{\cal J} F_{\cal L} | {\cal R},\Xi \rangle \,
\langle {\cal R},\Xi |H F_{\cal J}F_{\cal L} {\bf R}(\Omega)  | \Phi^{\pm}\rangle 
\label{sumstates}
\end{equation}
where the sum runs over all the possible spin-isospin components of
the $A$ particle system and the integral is extended over all the
spatial degrees of freedom of the system. The present scheme is valid
for the calculation of the expectation value of any operator that
commutes with the rotation operator.

We start from the spin and isospin sum. 
Let us focus on $\langle\Phi^{\pm}|F_{\cal L}|{\cal R},\Xi\rangle$
(here we do not include the factor $F_{\cal J}$ since it is a scalar in the
spin--isospin subspace). 
This factor is not zero only in those cases in which the action of
the operators in $F_{\cal L}$ over $|{\cal R},\Xi\rangle$ leads to the same
spin--isospin configuration as in $\Phi^{\pm}$.  If we designate by
$(0,0,0,0)$ the state saturated in spin--isospin and by
$(n_1,n_2,n_3,n_4)$ those states different from $(0,0,0,0)$ in $n_1$
particles with spin and isospin up, $n_2$ particles with spin up and
isospin down, $n_3$ particles with spin down and isospin up and $n_4$
particles with spin and isospin down, then the only $|{\cal
R},\Xi\rangle$ states that contribute in equation (\ref{sumstates})
are $(1,-1,-1,1)$ and $(-1,1,1,-1)$ as well as $(0,0,0,0)$.
Therefore the sum over the spin-isospin coordinates is reduced to only
three spin--isospin configurations of the $A$ nucleons.

The spatial integration is carried out by using the VMC in the same way as 
in the case of state independent correlations.
For the Metropolis random walk  we have used the following probability
distribution function
\begin{equation}
\omega({\cal R},\Omega)=
|F_{\cal J}({\cal R})F_{\cal L}^c({\cal R})|^2
|\langle\Phi^{\pm}|{\cal R},\Xi_{1}\rangle 
\langle{\cal R},\Xi_{1}|{\bf R}(\Omega)|\Phi^{\pm}\rangle |
\end{equation}
where $F_{\cal L}^c$ is the central part of the linear state dependent
factor $F_{\cal L}$ and  $\Xi_1$ stands for the spin and isospin configuration
of the nucleons equal to that of the state under description.
This function has been used in previous works providing a good performance.

In order to obtain the optimum set of variational parameters in the wave
function we have worked as follows. First we have optimized the
state--independent trial wave function. The search of the optimum set
of non linear parameters has been performed by means of the simplex
algorithm. Once this has been accomplished the state dependent trial
wave function is built by using the same Jastrow and model functions.
For the non linear parameters in the $g^{(k)}(r)$ functions of
equation (\ref{expansion}) we use the same ones as in the Jastrow function
$f(r)$. The only new variational parameters are the $a_m^{(k)}$ that
are fixed by solving the generalized eigenvalue problem.  The use of
the same set of non linear parameters does not lead to any appreciable
loss of accuracy, but it conveys  to a substantial reduction of the
computing time because the optimization of the non--linear parameters
for each trial wave function, as it was done in previous works, is
very time consuming.  This reduction is specially convenient for
heavy nuclei.

\section{Results}
\label{sec.results}

We have explored three possibilities of the correlation factor and the
two models for the intrinsic function.
For the correlation factor the three parameterizations considered are:
i) a central Jastrow factor only, J; 
ii) a linear state dependent correlation factor, LO; 
and iii) a central Jastrow times a linear state dependent, JLO.  
The JLO is the most general form of the correlation factor used here.
For the model part we have
used either a shell model built from Harmonic Oscillator orbitals
(spherical, HO, or deformed, HOd), 
or a function based on the Margenau-Brink model, MB.
For $^{12}$C and $^{16}$O, and within the MB model, 
we have also considered a linear geometry for the alpha clusters, describing
a different state with symmetry $D_{\infty h}$ and angular momentum and
parity given in table \ref{table1}.
These states have been also described by means of the Harmonic Oscillator
shell model by using the orbitals obtained in the limit of zero
intercluster distance given in table \ref{table1}.
The model wave function for these states is denoted by HO-l and MB-l when
using the Harmonic Oscillator and the Margenau-Brink models, respectively.

\subsection{Correlation factor}

\begin{figure}
\begin{center}
\epsfbox{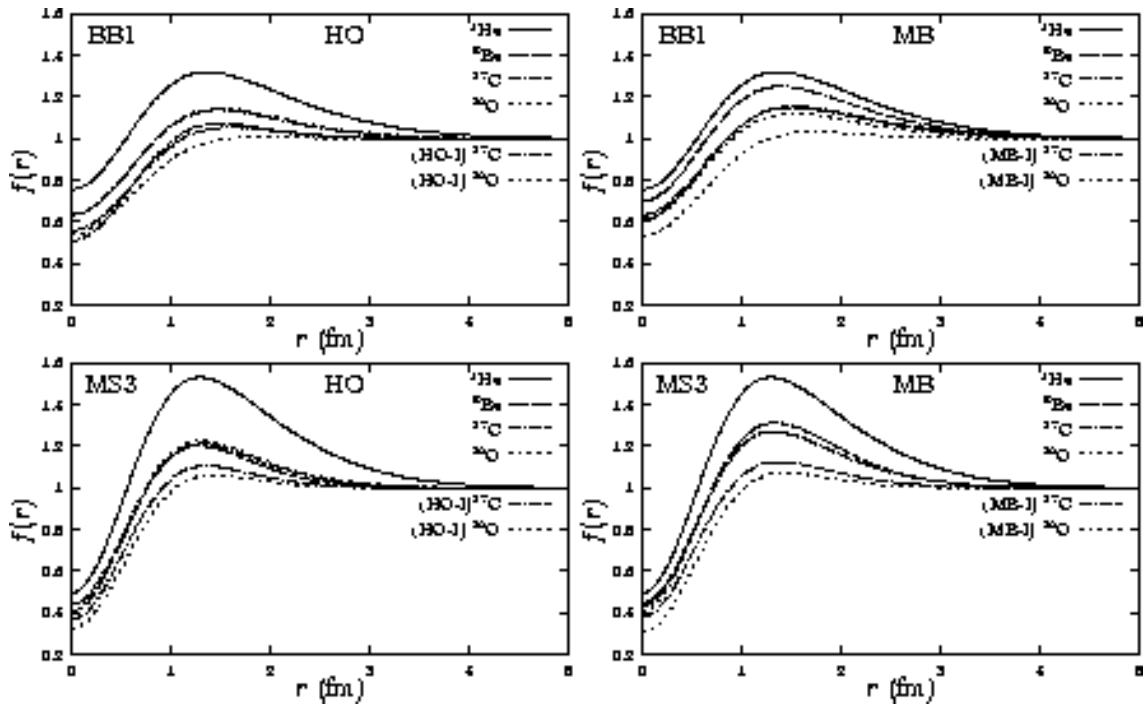}
\end{center}
\caption{
\label{figure1} 
Jastrow functions for the two potentials and the different trial
wave functions studied.
The upper plots correspond to the BB1 interaction and the lower
plots to the MS3 one.
The left hand side figures correspond to HO model wave functions and
the right hand side ones to Margenau-Brink model wave functions.
}
\end{figure}

In figure \ref{figure1} we show the optimal correlation function of
the Jastrow factor for both interactions and model wave functions.  In
general, the qualitative behaviour is the same for all of the cases
although quantitatively the function depends on both the model
function and the potential.  At short distances these functions are
mainly governed by the short range part of the potential. The depth of
the minimum at the origin is mostly determined by the strength of the
core of the interaction, that is higher in the MS3 than in BB1.  At
medium distances, $r\approx$ 1.2 fm, the correlation function has a
maximum.  The value of this maximum depends on the potential used, on
the model function and on the state of the nucleus under study.  For
the ground state, the height of the maximum decreases with the number
of nucleons for both models.  For the linear configurations of
$^{12}$C and $^{16}$O, the maximum is bigger than in the corresponding
ground state, and tends to the same values obtained for $^8$Be 
and $^4$He, this is because the clustering
is more accused in the linear geometry.  This is an example of the
interplay between the model function and the correlation factor.  It
is worth mentioning here that the energy is not very sensitive to this
part of the correlation function, and it is the short range part the
ultimate responsible of the binding.

\begin{figure}
\begin{center}
\epsfbox{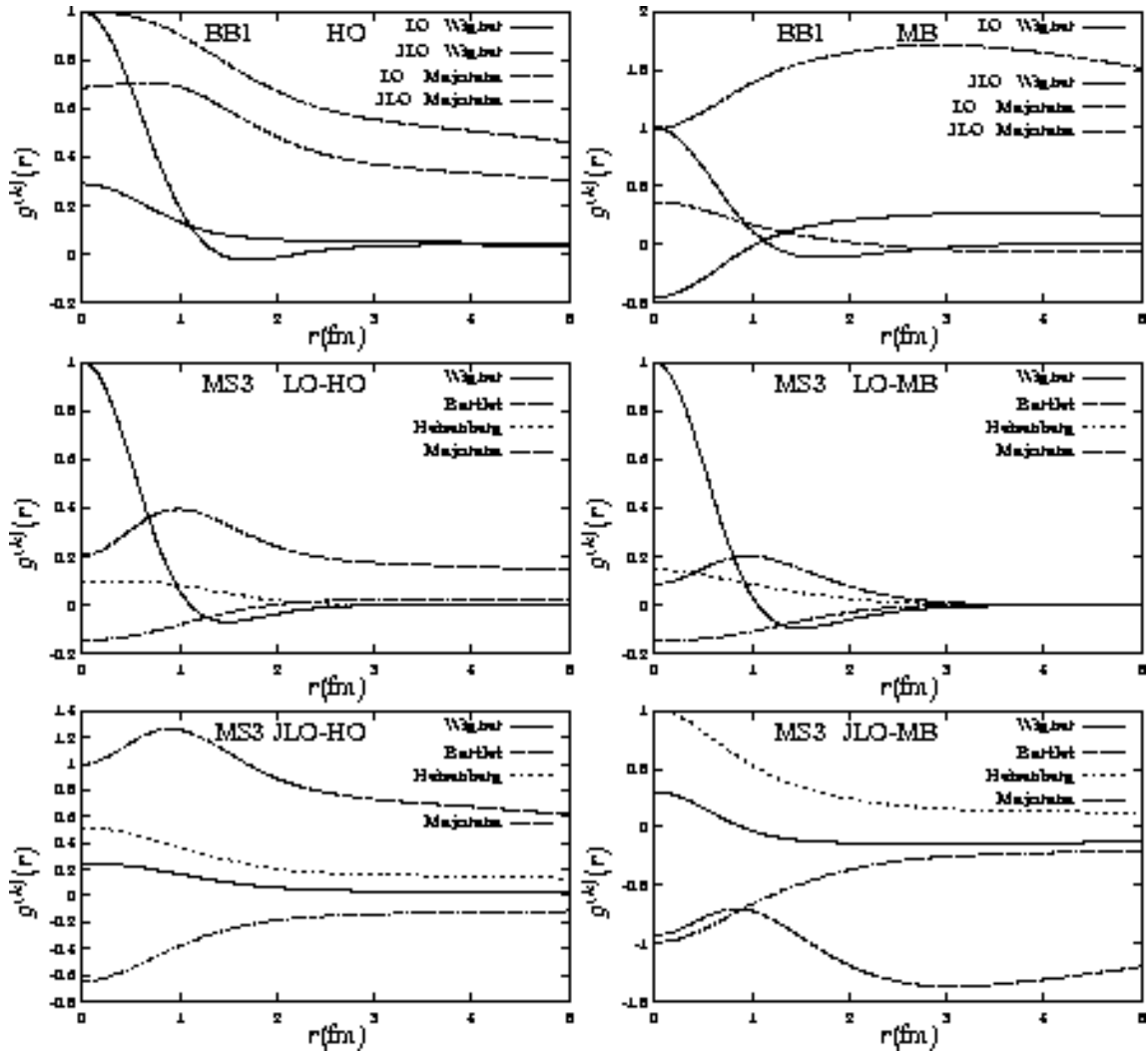}
\end{center}
\caption{
\label{figure2} 
Correlation functions of the linear correlation factor corresponding
to both the LO and JLO trial wave functions for the ground
state of the $^{12}$C nucleus. The upper panels correspond to the
BB1 interaction and the others to the MS3 force.
}
\end{figure}

The correlation functions of the linear correlation factor show, as
it could be expected, different forms in different channels.
This is due to the different
radial dependence of the potential on the different channels.  In
figure \ref{figure2} we plot the correlation functions obtained for 
the ground state of $^{12}$C in the LO and JLO approximations for the two
interactions and model functions considered in this work.  This
nucleus is representative of the different nuclei studied here.  Note
that the BB1 potential only has Wigner and Majorana channels.  Within
the LO approximation and for both model wave functions, 
the correlation functions of the Wigner and
Majorana channels present the same qualitative behaviour for both
interactions.  For the MS3 potential the correlation functions of the
Bartlett and Heisenberg channels are practically the same in magnitude but
with opposite sign.  In principle one could expect that both functions
are the same, except for the sign, because of the symmetry of the
interaction in these channels.  The reason of the difference is not
the parameterized solution used in this work because similar
differences have been found previously by using a different
methodology \cite{guardiola96} with the same type of wave function.

The inclusion of the Jastrow factor modifies the correlation functions
of the different channels.  The effect is more accused in the case of
the Wigner channel, where the correlation function practically vanishes.
For the other channels, the qualitative structure is similar to that
obtained in the LO approach, with the only exception of the Majorana
channel with Margenau-Brink model functions for both interactions. 

\subsection{Energy}

\fulltable{
\label{table2}
Binding energy and root mean square radius for the lowest
energy states of a given spatial arrangement calculated with the 
BB1 and MS3 interactions and the two model functions considered.
The variational parameters of the model functions are also reported.
Distances are given in fm and energies in MeV and $\beta$ in fm$^{-1}$.
In parentheses we show the statistical error.
}
\br
& \multicolumn{7}{c}{BB1} &\\
\mr
Nucleus & \multicolumn{2}{c}{wavefunction} 
&  \multicolumn{2}{c}{J} & \multicolumn{2}{c}{LO} &\multicolumn{2}{c}{JLO} \\
\crule{1} & \crule{2} & \crule{2}& \crule{2}& \crule{2} \\
& Model & $\beta;~d$ 
& $E$ & $\langle r^2\rangle^{1/2}$ 
& $E$ & $\langle r^2\rangle^{1/2}$ 
& $E$ & $\langle r^2\rangle^{1/2}$\\
\mr
$^4$He & HO & 0.68;1.   & -37.93(2) & 1.39(2) 
& -37.31(2) & 1.41(2) & 38.00(2) & 1.39(2) \\
\mr
$^8$Be & HOd & 0.76;0.61 
& -64.41(3) & 2.32(4) & -69.66(3) & 2.36(5) & -71.34(3) & 2.38(5) \\
       & MB  & 0.75;3.95 
& -73.65(2) & 2.35(3) & -72.59(3) & 2.41(4) & -74.86(3)  & 2.30(4)\\
\mr
$^{12}$C & HOd & 0.55:1.40 
& -101.06(4) & 2.39(4) & -109.72(8) & 2.40(6) & -114.20(7) & 2.39(6) \\
         & MB  & 0.74;3.87 
& -112.26(4) & 2.50(4) & -109.75(6) & 2.64(7) & -117.64(8) & 2.44(6) \\
\mr
$^{16}$O & HO  & 0.64;1.  
& -151.84(3) & 2.36(2) & -159.2(2) & 2.30(5) & -171.0(1) & 2.34(5) \\
         & MB  &  0.74;2.67 
& -166.56(5) & 2.37(3) & -170.5(2) & 2.35(5) & -180.7(1) & 2.33(5) \\
\mr
$^{12}$C & HO-l  & 0.69;0.56 
& -90.54(3)  & 3.19(4) & -96.80(6)  & 3.33(8) & -101.70(6) & 3.27(7) \\
         & MB-l  & 0.64;3.82 
& -100.48(3) & 3.40(3) & -99.77(4) & 3.55(8) & -107.62(5) & 3.37(5)\\
\mr
$^{16}$O & HO-l  & 0.66;0.52 
& -115.84(3) & 4.12(4) & -121.86(9) & 4.29(8) & -130.6(1) & 4.2(1)\\
         & MB-l  &  0.70;3.42 
& -127.19(3) & 4.12(3) & -128.7(1) & 4.17(8) & -135.96(9) & 4.09(8)\\
\br
& \multicolumn{7}{c}{MS3} &\\
\mr
Nucleus & \multicolumn{2}{c}{wavefunction} 
&  \multicolumn{2}{c}{J} & \multicolumn{2}{c}{LO} &\multicolumn{2}{c}{JLO} \\
\crule{1} & \crule{2} & \crule{2}& \crule{2}& \crule{2} \\
& Model & $\beta;~d$ 
& $E$ & $\langle r^2\rangle^{1/2}$ 
& $E$ & $\langle r^2\rangle^{1/2}$ 
& $E$ & $\langle r^2\rangle^{1/2}$\\
\mr
$^4$He & HO  & 0.61;1.   
& -27.11(3) & 1.43(3) & -26.03(3) & 1.52(3) & 30.17(3) & 1.40(2)\\
\mr
$^8$Be & HOd & 0.67;0.65 
& -43.09(4) & 2.42(4) & -46.2(1) & 2.51(5) & -54.07(6) & 2.45(5)\\
       & MB  & 0.66;4.43 
& -50.55(4) & 2.57(4) & -47.9(2) & 2.70(7)  & -57.6(2) & 2.51(9)\\
\mr
$^{12}$C & HOd & 0.54;1.42 
& -66.44(5) & 2.35(4) & -69.7(2) & 2.41(6) & -83.6(1)  & 2.36(6)\\
         & MB  & 0.70;3.58 
& -72.72(5) & 2.55(4) & -69.8(4)  & 2.62(9) & -87.3(2)  & 2.47(7)\\
\mr
$^{16}$O & HO  & 0.62;1.   
& -104.20(5) & 2.32(2) & -102.1(4)   & 2.34(6) & -128.4(2)   & 2.30(7)\\
         & MB  & 0.71;2.74 
& -112.50(6) & 2.40(3) & -101.5(4) & 2.43(6) & -134.3(2) & 2.34(6)\\
\mr
$^{12}$C & HO-l  & 0.66;0.58 
& -57.85(4) & 3.20(4) & -58.9(2) & 3.4(1) & -73.7(2) & 3.26(9)\\
         & MB-l  &  0.64;3.96 
& -65.77(4) & 3.54(3) & -60.5(1) & 3.66(9) & -77.8(1) & 3.5(1)\\
\mr
$^{16}$O & HO-l  & 0.63;0.54 
& -72.49(5) & 4.13(5) &  -69.1(4) & 4.3(2) & -93.3(2) & 4.2(2) \\
         & MB-l  & 0.63;3.89 
& -81.36(4) & 4.60(4) &  -70.8(4) & 4.7(1) & -98.0(4) & 4.6(2) \\ 
\br
\endfulltable

In table \ref{table2} we report the total energy and the root mean
square radius for the lowest energy states of the nuclei studied in
this work obtained from different trial wave functions and working
with the two potential considered here.  The wave functions have been
obtained by projecting to $K=0$ and $J=0$ the configurations proposed
for both the ground state and the states built from the linear
geometry for $^{12}$C and $^{16}$O.  The values of the variational
parameters in the model function are also shown in the table for each
nucleus, $\beta$ is the oscillator parameter in both the HO and MB
approaches and $d$ is either the deformation in the HOd model or the
intercluster distance in the MB model.

The harmonic oscillator parameter $\beta$ is larger with the BB1
interaction than with the MS3 one.  Therefore BB1 gives rise to a
bigger confinement of the nucleons as compared with MS3.  This can be
also concluded from the values of the root mean square radius. It is
worth to point out here that this quantity and, in general, those
related to the one body spatial density are mainly governed by the
model part of the wave function.  On the other hand, the deformation
for HO and intercluster distances for MB, are similar for both
interactions.  This can be understood by considering that the major
differences between the potentials take place at short distances, that
are accounted by the Jastrow factor, and to the fact that at medium
and large distances both forces are similar and the behaviour is
mainly governed by the model part of the wave function.  With respect
to the binding energies, they are bigger when the BB1 potential is
used.  This is because this potential was fitted by using non
correlated wave function, giving rise to an over binding when a more
realistic model is used.  In spite of this, the trend is the same for
all of the states considered and trial wave functions studied; for
example the MB model wave functions always constitute a better
variational option than the HO one.  In general, within the LO
model, the differences between MB and HO are very small for both
interactions.

\begin{figure}
\begin{center}
\epsfbox{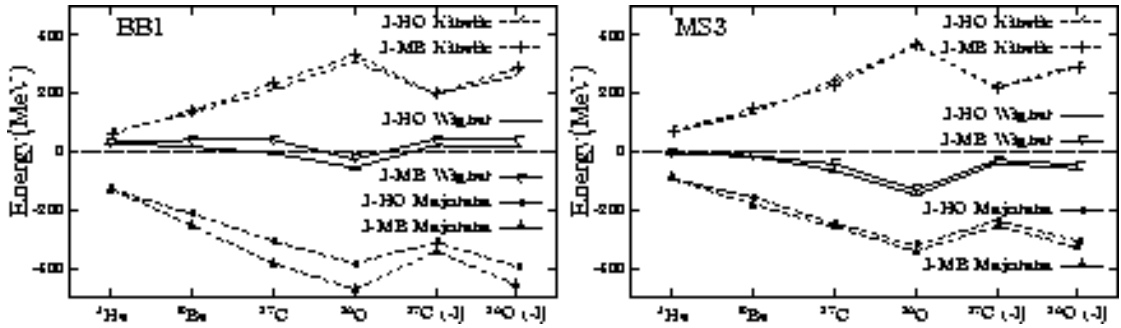}
\end{center}
\caption{
\label{figure3} 
Contribution to the total energy of the kinetic energy and the
channels of the nuclear potential for all of the nuclei
here studied including only central Jastrow correlations.
The left hand side plot corresponds to the results obtained with the BB1
potential and the right hand side plot to the MS3 interaction.
The symbol (-l) stands for the linear geometry of the alpha clusters
in the MB model and for the corresponding limiting harmonic oscillator
orbitals in the HO model.
The lines are for guiding the eyes.
}
\end{figure}
In order to get a deeper insight on the different correlation mechanisms
included in the variational wave functions 
we have analysed the different contributions to the
total energy.  
In figure \ref{figure3} we plot the contribution to the total 
energy of the kinetic energy as well as the energy from the
Wigner and Majorana channels for the ground state of the nuclei here studied
and the linear configurations of the $^{12}$C and $^{16}$O nuclei.  
This is done for the two potentials and model wave functions considered and
including only central Jastrow type correlations, J.
For both potentials, the HO functions show larger binding in the Wigner channel
than the MB ones while the opposite holds for the Majorana channel. 
The differences in the Majorana channel are bigger than in the Wigner one 
giving rise to the lower energy provided by the Margenau-Brink model.
The expectation values of the Bartlett and
Heisenberg channels calculated from state independent wave
functions have the same value with opposite sign, and therefore they
cancel out.  

The Jastrow factor, J, and the linear operatorial correlations, LO,  involve 
dynamical mechanisms of different nature, and therefore the origin of
the binding energy obtained with them can be very different.
A hint on the different correlation mechanisms included by the J and LO factors
is that, when both are taken into account simultaneously, 
a significant increase of the binding energy is obtained, specially with 
the MS3 interaction.  
\begin{figure}
\begin{center}
\epsfbox{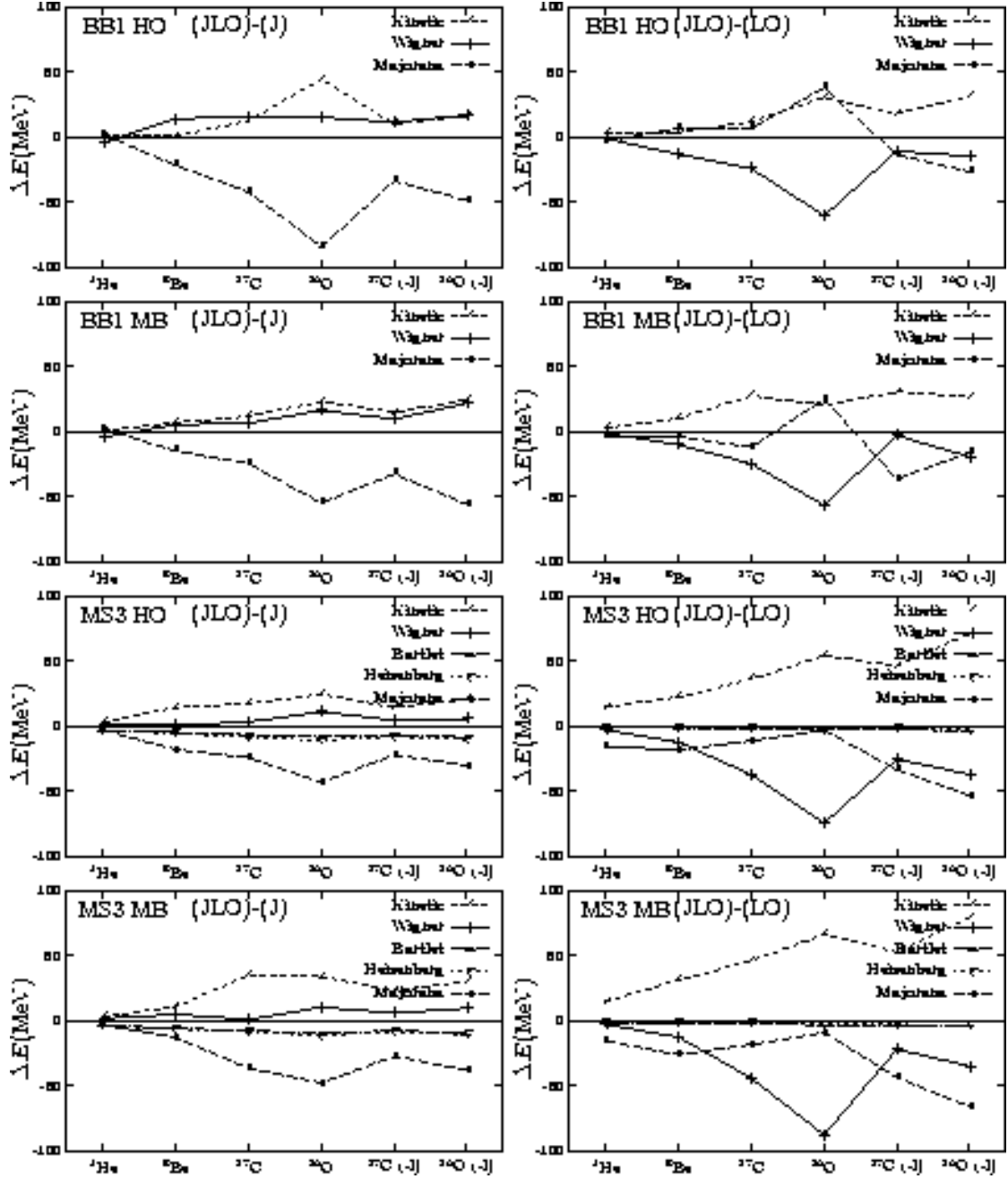}
\end{center}
\caption{
\label{figure4} 
Difference of the kinetic energy and the
channels of the nuclear potential between the values obtained
form the J and LO wave functions and those calculated from the JLO
trial wave function.
The upper plots correspond to the BB1
potential and the lower plots to the MS3 interaction.
The symbol (-l) stands for harmonic oscillator orbitals of
the linear geometry.
The lines are for guiding the eyes.
}
\end{figure}

To further elucidate on the effects of the different correlation
mechanisms on the binding energy, we plot in figure \ref{figure4} the
differences between several expectation values related to the energy
calculated from different wave functions.
We show the differences between the expectation values of
the kinetic energy and all of the channels of the nuclear potential
calculated from the J and the JLO models ((JLO)-(J)) and between the
LO and JLO models ((JLO)-(LO)). The former can be interpreted as the effects
of the linear state dependent correlations on these quantities while the
latter are the effects of the central Jastrow correlations.
When these differences take positive values, the action of the corresponding
operator is to reduce the binding and the opposite holds for negative values.

The effect of both state-dependent and Jastrow correlations is to
increase the expectation value of the kinetic energy, reducing the
binding.  This effect is more accused when the MS3 interaction is
employed.  This is also the case of the Wigner channel when operators
are included.  An exception to this is the $^4$He nuclei with the BB1
interaction that shows the opposite trend.  The effect of the Jastrow
correlations on the Wigner channel is to increase the binding energy.
The effect of the state dependent correlations is more accused for the
BB1 potential and the effect of the Jastrow factor is more important
for the MS3 interaction. In the latter case, it is an order of
magnitude bigger when using MS3 than for BB1 for the ground states and
it is reduced for the other states.  In the Majorana channel,
operators tend to increase the binding energy except for $^4$He
with the BB1 interaction.  In the case of Jastrow correlations,
the effects depend on the potential.  Thus when using the MS3
interaction, the tendency is to increase the binding energy. However,
for the BB1 potential, it depends on the nucleus and on the model wave
function.  Finally the effects on the Bartlett and Heisenberg channels,
that appear only in the MS3 interaction, are very similar. For these
channels, both the Jastrow and the operators tend to increase the
binding energy. The effect is more important, by a factor 3-4, for the
operators than for the Jastrow.

Therefore the action of the state dependent correlations is to
increase the binding by means of the Majorana channel. In the case of
the BB1 interaction, the reduction of the energy coming from this
channel is bigger than the enhancement coming from the sum of the Wigner
channel and the kinetic energy.  In the case of the MS3 potential, the
increasing in the binding energy given by the Majorana channel is
roughly the same as the decreasing provided by the sum of the Wigner
channel and the kinetic energy. As a consequence, the gain in the
total binding energy due to the operators, is the sum of the Bartlett and
Heisenberg channels.  The $^4$He nuclei is an exception. In the case
of the BB1 interaction, the gain comes from the Wigner channel and it
is very small (0.1 MeV).  For the MS3 potential, the Majorana is not
enough to compensate for the decrease in the binding energy provided
by both the Wigner energy and the kinetic energy.

The action of the Jastrow correlation factor is to increase the binding
energy by means of the Wigner channel. For the BB1 potential,
the effect on Majorana channel depend on the nucleus. This is not the
case of the MS3 interaction, for which the effect of the Jastrow on the
Majorana channel is to increase the binding energy.

\fulltable{
\label{table3}
Excitation energies (in MeV) with respect to the lowest energy state of
a given configuration for the nuclei here studied obtained from the
different trial wave functions and the two potentials considered.  In
parentheses we show the statistical error.  }  
\br & & \multicolumn{4}{c}{BB1} &\\ 
\mr & $J^\pi$ & HO/MB & J(HO/JMB) &
LO(HO/MB) & JLO(HO/MB) & Exp. \\ 
\mr $^8$Be & $2^+$ & 3.24(3)/3.42(3)
& 3.44(3)/3.73(3) & 3.80(4)/3.71(3) & 3.90(4)/3.91(3) & 3.040\\ 
& $4^+$ & 11.72(4)/11.76(4) & 12.40(5)/12.89(4) & 13.26(6)/12.63(5) &
13.48(5)/13.59(5) & 11.400\\ 
\mr $^{12}$C & $2^+$ & 2.97(4)/3.37(4) &
3.11(4)/3.83(4) & 3.61(8)/3.94(6) & 3.65(6)/4.10(7) & 4.439\\ 
& $4^+$ & 11.5(1)/12.24(6) & 12.14(9)/13.71(6) & 13.6(2)/13.7(1) &
13.8(2)/14.7(1) & 14.083\\ 
& $0^+$ & 6.01(8)/10.08(7) &
12.52(7)/11.76(7) & 13.9(1)/10.0(1) & 12.5(1)/10.0(1) & 7.654\\ 
& $3^-$ & -/6.70(8) & -/7.8(1) & -/6.6(2) & -/7.7(2) & 9.641\\ 
\mr $^{16}$O & $3^-$ & -/- & -/16.3(1) & -/15.4(5) & -/15.9(4) & 11.600\\
\br & & \multicolumn{4}{c}{MS3} &\\ \mr $^8$Be & $2^+$ &
2.23(4)/2.65(4) & 2.95(4)/3.18(4) & 3.1(1)/3.1(1) & 3.46(6)/3.5(1) &
\\ 
& $4^+$ & 8.32(8)/9.11(6) & 10.71(7)/10.80(7) & 11.0(2)/10.4(2) &
11.9(1)/12.0(3) & \\ 
\mr $^{12}$C & $2^+$ & 2.41(6)/2.97(6) &
3.23(5)/2.42(4) & 3.4(2)/3.7(4) & 3.6(1)/3.5(2) & \\ 
& $4^+$ & 9.9(1)/10.2(1) & 12.6(1)/12.6(1) & 12.7(4)/12.2(9) 
& 14.0(3)/13.1(4) & \\ 
& $0^+$ & -/- & 8.59(9)/6.95(9) & 10.8(4)/9.3(5) & 9.9(3)/9.5(3) & \\ 
& $3^-$ & -/- & -/7.7(1) & -/6.6(9) & -/8.4(4) & \\ 
\mr $^{16}$O & $3^-$ & -/- & -/15.2(2) & -/10.5(1) & -/16.0(8) & \\ 
\br 
\endfulltable
The model used in this work provides the ground state and also some
other excited states by means of the projection operation of the total
angular momentum and parity.  In table \ref{table3} we show different
excitation energies with respect to their corresponding lowest
energy ones reported in table \ref{table2}. We give those states that
can be assigned to some experimental nuclear excited states.  The
excitation energy is almost independent of the model wave function and
the effect of state dependent correlations is to increase slightly the
value of the excitation energy.  In general it can be concluded that
the simple models used here and the physical picture behind them
provide a reasonable description of these nuclear states.  \\

\subsection{One- and two- body densities}

To better understand the differences between the two model wave
functions used in this work we study both the single particle and the
two--body densities.  We shall consider here the spherical average of
these two functions normalized to unity which are defined as 
\begin{eqnarray}
\rho^{(1)}(r)=\int d\tau |\Psi(\tau)|^2\left\{\frac{1}{A}
\sum_{i=1}^A\frac{1}{r^2}\delta(r-|\vec{r_i}-\vec{R}|)\right\}
\\
\rho^{(2)}(r_{12})=\int d\tau |\Psi(\tau)|^2\left\{\frac{2}{A(A-1)}
\sum_{i<j}^A\frac{1}{r_{12}^2}\delta(r_{12}-|\vec{r_i}-\vec{r}_j|)\right\}
\end{eqnarray}
where $\tau$ stands for all of the particles' spatial coordinates and
intrinsic degrees of freedom, and
$\vec{R}=\frac{1}{A}\sum_{i=1}^A\vec{r}_i$ is the center 
of mass coordinate.
\begin{figure}
\begin{center}
\epsfbox{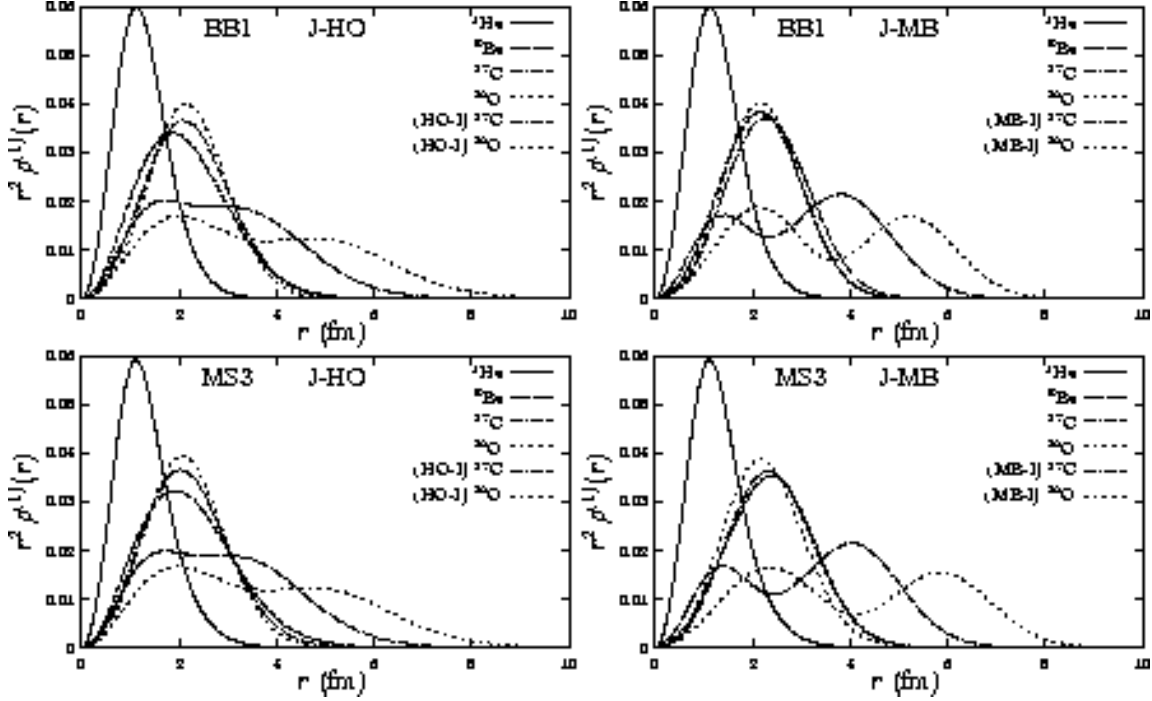}
\end{center}
\caption{
\label{figure5} 
One-body radial density for all of the states
here studied calculated from a trial wave function with a central
Jastrow correlation factor.  The upper plots correspond to the BB1
interaction and the lower plots to the MS3 one.  The left hand side
figures correspond to HO model wave functions and the right hand side
ones to MB model wave functions, both of them with a Jastrow factor.
}
\end{figure}
The one-- body density is the probability distribution of the nucleons
with respect to the center of mass and the two-- body density is the
probability distribution with respect to a given nucleon.  In figure
\ref{figure5} we plot the one body radial density, $r^2\rho^{(1)}(r)$,
obtained from the Jastrow correlated wave functions for the
interactions and model wave functions analysed in this work.  For the
ground state of the different nuclei, the radial one body density has
one maximum whose height and width depend on both the nucleus and the
model wave function.  For the linear geometry in $^{12}$C and $^{16}$O
the situation is different, with two maxima for the two model wave
functions considered.  Within the Margenau-Brink model these maxima
are interpreted as the position of the alpha clusters.  Note that for
$^{12}$C the center of mass coincides with the center of the
central cluster, while for $^{16}$O the location of the $\alpha$-clusters is
symmetric with respect to the position of the center of mass.  When
the model part of the wave function is built from a HO shell model,
the position of the maxima is related to the extremes of the single
particle orbitals, given in table \ref{table1}.  The one body density
is mainly governed by the model function of the trial wave function.
The effect of the correlations is indirect through the change of the
variational parameters of the model function induced by the inclusion
of the correlations.

The radial density is very similar for all of the cases considered
except for $^{4}$He, that is of shorter range.  The use of different
interactions does not modify the structure of this function for the
ground state and only
changes slightly the position and height of the maximum, that are
reduced when the MS3 potential is used with respect to the BB1 case.
The use of a different model wave function, with the same potential,
also induces minor changes.  With respect to the linear geometry, we
have obtained an appreciable dependence on the model wave function,
and a nearly negligible dependence on the nuclear interaction.  
For the linear configuration, the clusterization can be clearly observed
on this density when using both model wave functions. 
Note that the formation of alpha clusters is included explicitly in the
MB model but not in the HO one.

\begin{figure}
\begin{center}
\epsfbox{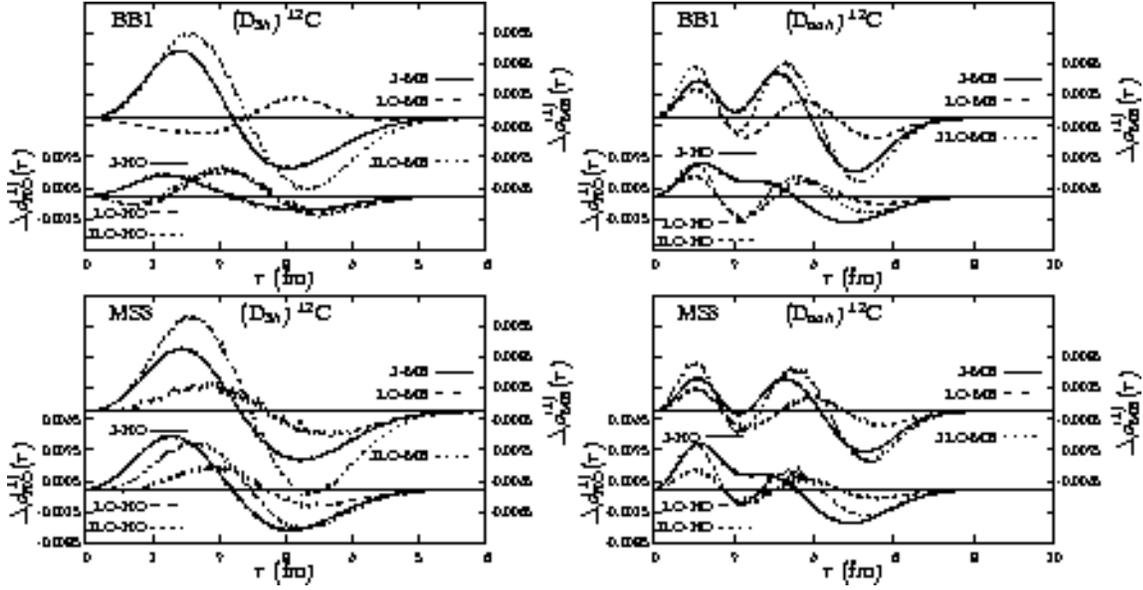}
\end{center}
\caption{
\label{figure6} 
One-body difference functions for the ground state configuration
and the linear configuration of $^{12}$C.  The upper plots correspond
to the BB1 interaction and the lower plots to the MS3 one.  The left
hand side figures correspond to the ground state and the right
hand side ones to the linear arrangement. For any figure the
upper (lower) part corresponds to MB (HO) model. The full,
dashed, and dotted lines correspond to J, LO, and JLO wave
functions, respectively.
}
\end{figure}

In order to analyse the effects of the different correlation mechanisms
we have calculated the difference between the densities obtained from a
correlated and an uncorrelated trial wave function. 
For the uncorrelated case we have used the same model wave function as
in the correlated trial wave functions
\begin{equation}
\Delta \rho^{(k)}_{\mu}(r) = 
r^2\left[\rho^{(k)}_{\mu}(r)-\rho^{(k)}_{\rm{uc}}(r)\right],~~~k=1,2.
\end{equation}
where $\rho^{(k)}_{\rm{uc}}(r)$ is the $k$-body density calculated
from the uncorrelated wave function and $\rho^{(k)}_{\mu}(r)$, where
$\mu$ stands for J, LO, JLO, corresponds to a density calculated from each
of these correlation factors.
In figure \ref{figure6} we plot these difference functions for $^{12}$C,
which is representative for all of the nuclei here studied. They have
been calculated with the two nuclear potentials and 
the two model wave functions here considered.
The effect of the central Jastrow correlations is to increase the density
at distances smaller than roughly 2 fm for the ground state and 4 fm
for the linear geometry, and to decrease this density for larger distances. 
For the wave functions built from HO
model functions and using the BB1 potential the effect of the
correlations is less important.  When only linear state dependent
correlations are included, LO, we have not found a systematic trend.
In the case of the linear geometry, correlations tend to increase the
density in the neighbourhood of the maxima and to reduce the density
around the minima as well as at distances greater than 4 fm, close to
the nuclear surface.  The only exception to this effect is for HO wave
functions with the MS3 interaction.

\begin{figure}
\begin{center}
\epsfbox{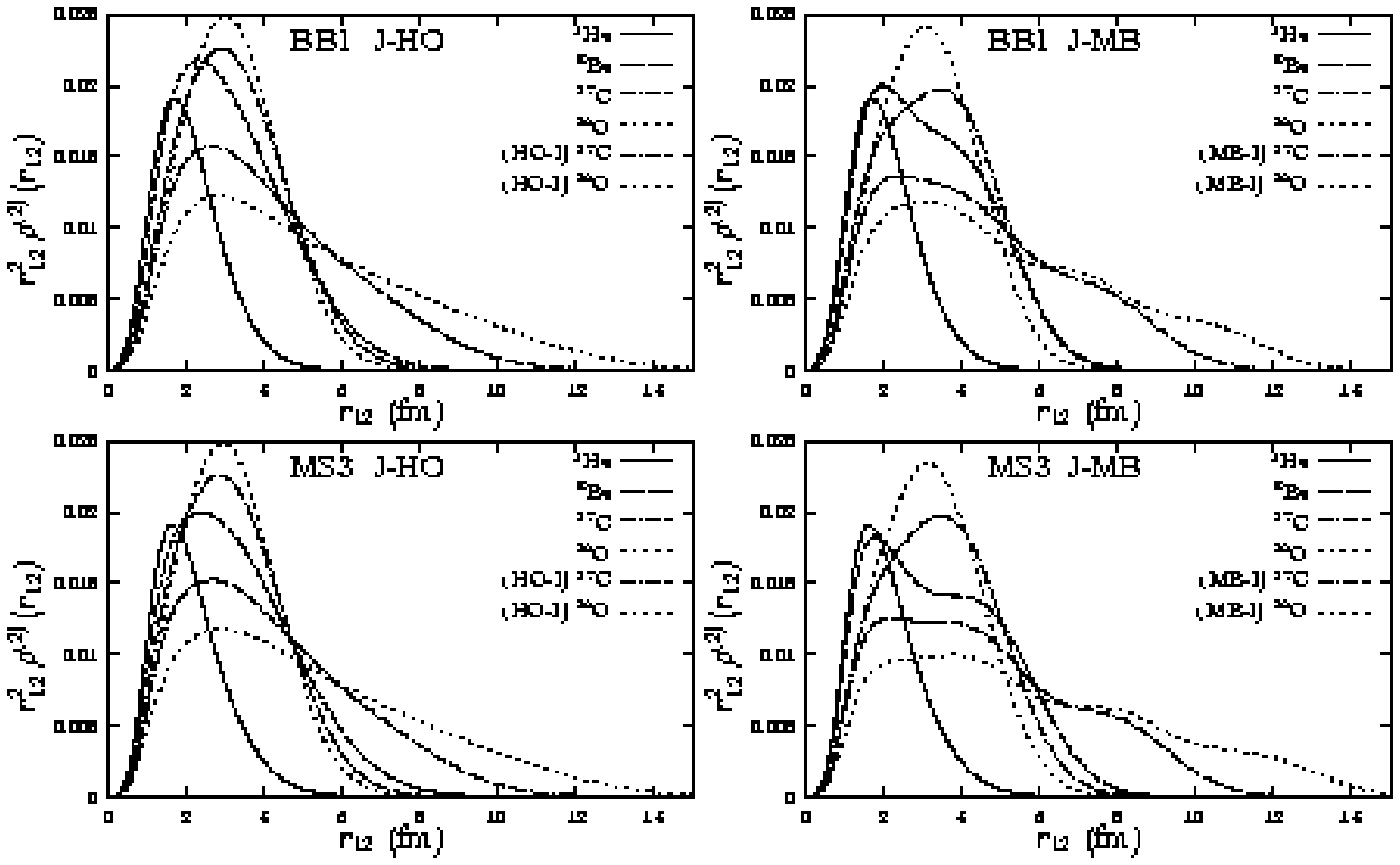}
\end{center}
\caption{
\label{figure7} 
Two-body radial density for all of the states here studied calculated
from a trial wave function with a central Jastrow correlation factor.
The upper plots correspond to the BB1 interaction and the lower
plots to the MS3 one.
The left hand side figures correspond to HO model wave functions and
the right hand side ones to Margenau-Brink model wave functions.
}
\end{figure}

In figure \ref{figure7} we plot the two body radial density obtained from
the Jastrow correlated wave function for the two different
interactions and model wave functions considered.
As can be seen, when HO wave
functions are employed, the results are roughly independent of the
nuclear potential and a greater dependence on the potential  is found
in the case of the MB wave functions.
The cluster structure of the nuclei here studied can be seen
on this density.  The different maxima correspond to interparticle
distances of nucleons in the same or in different clusters.  This can
be clearly noticed in $^8$Be and in the linear configuration of
$^{12}$C and $^{16}$O.  The ground state two-body density of $^{12}$C
and $^{16}$O presents only one maximum, i.e. the two characteristic
internucleon distances are very close.
This is because, in the ground state, the clusters are not very 
far from each other.  
On the other hand one does not have any particular nucleon grouping 
when the HO wave function is used and therefore there is only one 
maximum which becomes higher and narrower as the number of nucleons 
increases.  

\begin{figure}
\begin{center}
\epsfbox{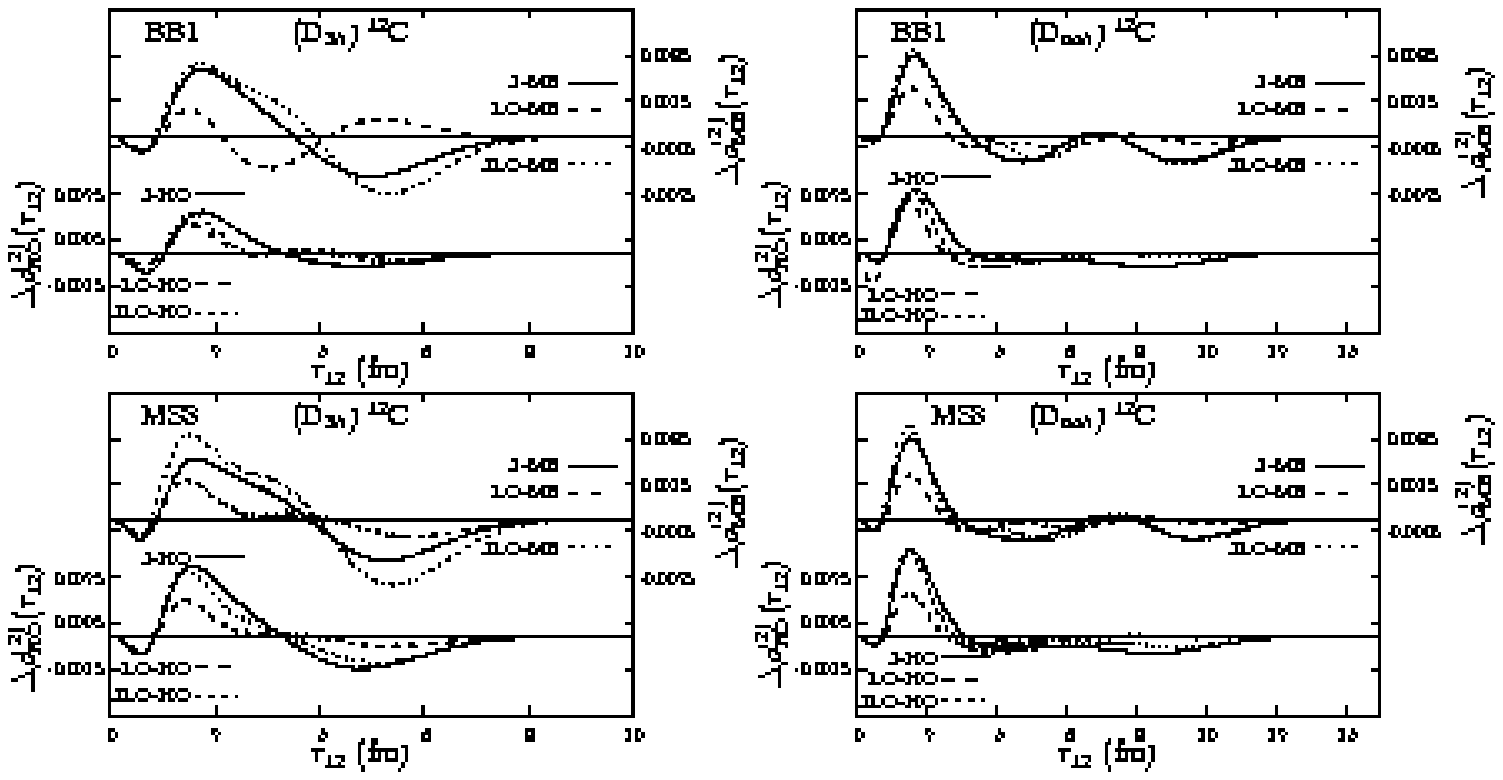}
\end{center}
\caption{
\label{figure8} 
Two-body difference functions for the ground state configuration and
the linear configuration of $^{12}$C.
The upper plots correspond to the BB1 interaction and the lower
plots to the MS3 one.
The left hand side figures correspond to the ground state and
and the right hand side ones to the linear arrangement.
For any figure the
upper (lower) part corresponds to MB (HO) model. The full,
dashed, and dotted lines correspond to J, LO, and JLO wave
functions, respectively.
}
\end{figure}
In figure \ref{figure8} we plot the functions $\Delta
\rho^{(2)}_{\mu}(r)$ with $\mu=$J, LO, JLO for $^{12}$C calculated
from the two different models and interactions here considered.  As it
was the case for the one body density, we have obtained a stronger
influence of the model wave function than of the nuclear potential on
these difference functions.  Two common features to all of the
difference functions have been found.  The first one is a negative
region at short inter-nucleon distances with a minimum located at around
0.5 fm.  This structure is generated by the decreasing of the density
around a given nucleon in order to avoid the repulsive core of the
potential.  The second element is a first maximum with a well defined
structure in the case of the linear configuration and a little bit
more diffuse in the ground state.  This maximum is a consequence of
the increase of the relative density at distances in the neighbourhood
of the minimum of the nuclear potential.  At larger nucleon-nucleon
distances the effects of the correlations, specially in the linear
configuration, are smaller.  It can be stated that within the
Margenau-Brink model correlations are focused on nucleons of the same
cluster, while that within the HO shell model, the main effects of
correlations take place between nucleons in orbitals with the same
spatial part.

\section{Conclusions}
\label{sec.conclusions}

A variational study of the ground and some bound states of the 
$^4$He, $^4$Be, $^{12}$C and  $^{16}$O nuclei starting from nucleon-nucleon
interactions is presented.
The binding energy, the root mean square radius and the one-- and 
two-- body densities are reported.
The calculations have been done by means of the Variational Monte Carlo 
method.

The variational trial wave function used consists of three factors including
several aspects of the nuclear dynamics induced by the nuclear interaction.
Short range correlations are accounted by a central Jastrow-type factor,
state dependent correlations are incorporated by a linear factor
depending on the
spin and isospin of the nucleons, and some medium and long range effects
by the model part of the wave function, that it is also antisymmetric.
The medium and long range effects considered here are the deformation of the
nuclear potential and the formation of alpha clusters.
Projection operators are used to obtain wave functions with the proper 
values of parity and total angular momentum.
This functional form has shown to be able to describe
several dynamic effects that give rise to different mechanisms
of lowering the energy depending on the nucleus, the state and the
potential.

The performance of the two model wave functions employed and the
relationship of the different elements of the variational wave
function have been studied.  The different correlation functions,
the size of the nucleus and the deformation
parameter or the intercluster distances have been obtained and compared for
the two different potentials employed.  The effects of the different
correlation mechanisms have been analysed by calculating the different
contributions to the total energy.  It has been found that the inclusion of
Jastrow and state dependent correlations increase the expectation
value of the kinetic energy. In the former case a substantial
reduction in the expectation value of the Wigner channel is obtained,
leading to the increase in the total binding energy that  is found
when these correlations are included. The other channels of the
potential, in general, also contribute to a bigger binding when
Jastrow correlations are included.  
In the case of state dependent correlations the effect on the Wigner
channel is the opposite, and it is the Majorana channel for the BB1
potential and the Majorana, Bartlett and Heisenberg channels for the
MS3 interaction, the responsible for the bigger binding energy.  When
using model wave functions including explicitly the alpha clustering
effect, lower values of the expectation value of the Majorana channel
are obtained, giving rise to the better performance of this ansatz from
a variational point of view. 

For the one body density, a small dependence on both the potential and the
model wave function for the ground state has been found
while the dependence on the model wave function is more accused
in the linear configuration.
The effects of the correlations are to increase the density at short
distances for the ground state and to make more accused the alpha
clustering for the linear configuration.  In the case of the two body
radial distribution a stronger dependence on the model wave function
has been found. This is due to the two typical internucleon distances
that appear in a Margenau-Brink type structure, one between nucleons
in the same alpha cluster and the second one between nucleons in
different clusters.  On this density, the effect of correlations is to
decrease the density around a given nucleon and increase it for
distances close to the minimum of the potential, avoiding the hard
core of the nuclear potential.  This is consistent with the short
range character assigned to the correlation factor, that  is mainly
focused on nucleons within the same alpha cluster in the case of
Margenau-Brink, and on nucleons with the same spatial part in the case
of HO model.

\ack
This work has been partially supported by the Ministerio de Ciencia y
Tecnolog\'{\i}a and FEDER under contract  BFM2002-00200,  
and by the Junta de Andaluc\'{\i}a.

\end{document}